\documentclass[preprint,floats,aps,epsfig,nofootinbib,superscriptaddress,amssymb,showpacs]{revtex4}
\usepackage{color}
\usepackage{graphicx}
\usepackage{tabularx}
\usepackage{amsmath}
\usepackage{amssymb}
\usepackage{datetime}
\usepackage[left]{lineno}
\usepackage{txfonts}
\usepackage{dcolumn}
\usepackage{bm}

\begin{document}


\title{Low energy neutrinos from stopped muons in the Earth}

\author{Wan-Lei Guo}
\email{guowl@ihep.ac.cn} \affiliation{Institute of High Energy
Physics, Chinese Academy of Sciences, P.O. Box 918, Beijing 100049,
China}

\begin{abstract}

We explore the low energy neutrinos from stopped cosmic ray muons in
the Earth. Based on the muon intensity at the sea level and the muon
energy loss rate, the depth distributions of stopped muons in the
rock and sea water can be derived. Then we estimate the $\mu^-$
decay and nuclear capture probabilities in the rock. Finally, we
calculate the low energy neutrino fluxes and find that they depend
heavily on the detector depth $d$. For $d = 1000$ m, the $\nu_e$,
$\bar{\nu}_e$, $\nu_\mu$ and $\bar{\nu}_\mu$ fluxes in the range of
13 MeV $ \leq E_\nu \leq$ 53 MeV are averagely  $10.8 \%$, $6.3\%$,
$3.7 \%$ and $6.2 \%$ of the corresponding atmospheric neutrino
fluxes, respectively. The above results will be increased by a
factor of 1.4 if the detector depth $d < 30$ m. In addition, we find
that most neutrinos come from the region within 200 km and the near
horizontal direction, and the $\bar{\nu}_e$ flux depends on the
local rock and water distributions.

\end{abstract}


\maketitle

\section{Introduction}

Atmospheric neutrinos are a very important neutrino source to study
the neutrino oscillation physics. In 1998, the Super-Kamiokande
(Super-K) experiment reported the first evidence of neutrino
oscillations based on a zenith angle dependent deficit of
atmospheric muon neutrinos \cite{Fukuda:1998mi}. Atmospheric
neutrinos are produced in the Earth's atmosphere as a result of
cosmic ray interactions and the weak decays of secondary mesons, in
particular pions and kaons \cite{Honda:2006qj}. At the same time, a
large amount of muons are also produced and some of them can
penetrate the rock and sea water of Earth's surface to significant
depths. These penetrating muons are the important background source
for some underground experiments \cite{Li:2014sea}. It is well known
that these muons will finally stop in the Earth and then produce the
low energy neutrinos through decay or nuclear capture
\cite{Measday:2001yr}. However, these neutrinos are not included in
the previous literatures \cite{Gaisser:1988ar, Honda:1995hz,
Battistoni:2005pd}. Here we shall focus on these neglected neutrinos
from stopped muons in the Earth.

Muons are the most numerous charged particles at sea level
\cite{PDG}. After losing energy by ionization and radiative
processes, the stopped $\mu^+$ in the rock and sea water will
generate two low energy neutrinos $\nu_e$ and $\bar{\nu}_\mu$
($E_\nu \leq 53$ MeV) through $\mu^+ \rightarrow e^+ + \nu_e +
\bar{\nu}_\mu$. Unlike $\mu^+$, the stopped $\mu^-$ may undergo
either decay or capture by the nucleus \cite{Measday:2001yr}. In the
nuclear capture case, a stopped $\mu^-$ can only produce a neutrino
$\nu_\mu$ with energy less than the muon mass. These low energy
neutrinos will be the background source in the searches of some
relevant physics, such as diffuse supernova relic neutrinos
\cite{Ando:2004hc,An:2015jdp}, dark matter annihilation in the Sun
\cite{Rott:2012qb} and our galaxy \cite{PalomaresRuiz:2007eu}, solar
neutrino conversion \cite{Collaboration:2011jza},  and proton decays
catalyzed by GUT monopoles in the Sun \cite{Ueno:2012md}. So it is
necessary for us to investigate the low energy neutrinos induced by
stopped muons in the Earth.

In this paper, we shall calculate the neutrino fluxes from stopped
cosmic ray muons in the Earth. In Sec. II, the stopped $\mu^\pm$
distributions in the rock and sea water will be given in terms of
the muon intensity at the sea level and the muon energy loss rate.
Sec. III is devoted to the $\bar{\nu}_e$ and $\nu_\mu$ energy
spectra from a stopped $\mu^-$. Based on the atomic capture and
nuclear capture abilities of 10 dominant elements in the upper
continental crust, we estimate the $\mu^-$ decay and nuclear capture
probabilities in the rock and sea water. In Sec. IV, we numerically
calculate the low energy neutrino fluxes according to the stopped
$\mu^\pm$ distributions and the $\mu^-$ decay probability, and
discuss their features. In addition, an approximation formula to
compute the neutrino fluxes has also been presented. Finally, our
conclusions will be given in Sec. V.


\section{Distributions of stopped muons}

Muons are the most numerous charged particles at sea level
\cite{PDG}. For the energy and angular distribution of cosmic ray
muons at the sea level, we use the following parameterization
\cite{Reyna:2006pv}
\begin{eqnarray}
I(p_\mu, \theta) = I_{\rm v}(\xi)\, \cos^3 \theta \;\; {\rm for}
\;\; p_\mu
> 1 {\rm GeV}\,, \label{Imu1}
\end{eqnarray}
with $\xi = p_\mu \cos \theta$ and $\theta$ is the zenith angle. The
vertical muon intensity is given by
\begin{eqnarray}
 I_{\rm v}(p_\mu) = c_1 p_\mu^{-(c_2 + c_3 \log p_\mu + c_4 \log^2 p_\mu  +c_5 \log^3 p_\mu
 )} \,,
\end{eqnarray}
where $c_1 = 0.00253 \, {\rm cm^{-2} s^{-1} sr^{-1} GeV^{-1}}$, $c_2
= 0.2455$, $c_3 = 1.288$, $c_4 = -0.2555$ and $c_5= 0.0209$. It is
found that Eq. (\ref{Imu1}) is valid for all zenith angles and the
muon momentum $p_\mu > 1$ GeV \cite{Reyna:2006pv}. For $p_\mu \leq
1$ GeV, the muon energy spectrum is almost flat and the
corresponding angular distribution is steeper than $\cos^2 \theta$
\cite{PDG}. Therefore we assume
\begin{eqnarray}
I(p_\mu , \theta) =  0.00389 \cos^3 \theta  \;\; {\rm for} \;\;
p_\mu \leq 1 {\rm GeV}\,,  \label{Imu2}
\end{eqnarray}
in the units of ${\rm cm^{-2} s^{-1} sr^{-1} GeV^{-1}}$. In terms of
of Eqs. (\ref{Imu1})-(\ref{Imu2}), the total muon flux for a
horizontal detector can be derived from
\begin{eqnarray}
J_\mu  = 2 \pi \int I(p_\mu, \theta) \cos \theta \; d\cos \theta \,
d p_\mu = 1 \, {\rm cm^{-2} \, min^{-1}} \;,
\end{eqnarray}
which is familiar to experimentalists for horizontal detectors
\cite{PDG}. The muons with $p_\mu > 1 {\rm GeV}$ make a $71\%$
contribution to $J_\mu$. With the help of the muon charge ratio in
Fig. 6 of Ref. \cite{Naumov:2002dm}, we calculate the $\mu^+$ flux
$J_{\mu^+} = 0.0093 \, {\rm cm^{-2} s^{-1}}$ and the $\mu^-$ flux
$J_{\mu^-} = 0.00737 \, {\rm cm^{-2} s^{-1}}$, and the corresponding
muon charge ratio $J_{\mu^+}/J_{\mu^-} = 1.26$.

\begin{figure}[htb]
\begin{center}
\includegraphics[scale=0.38]{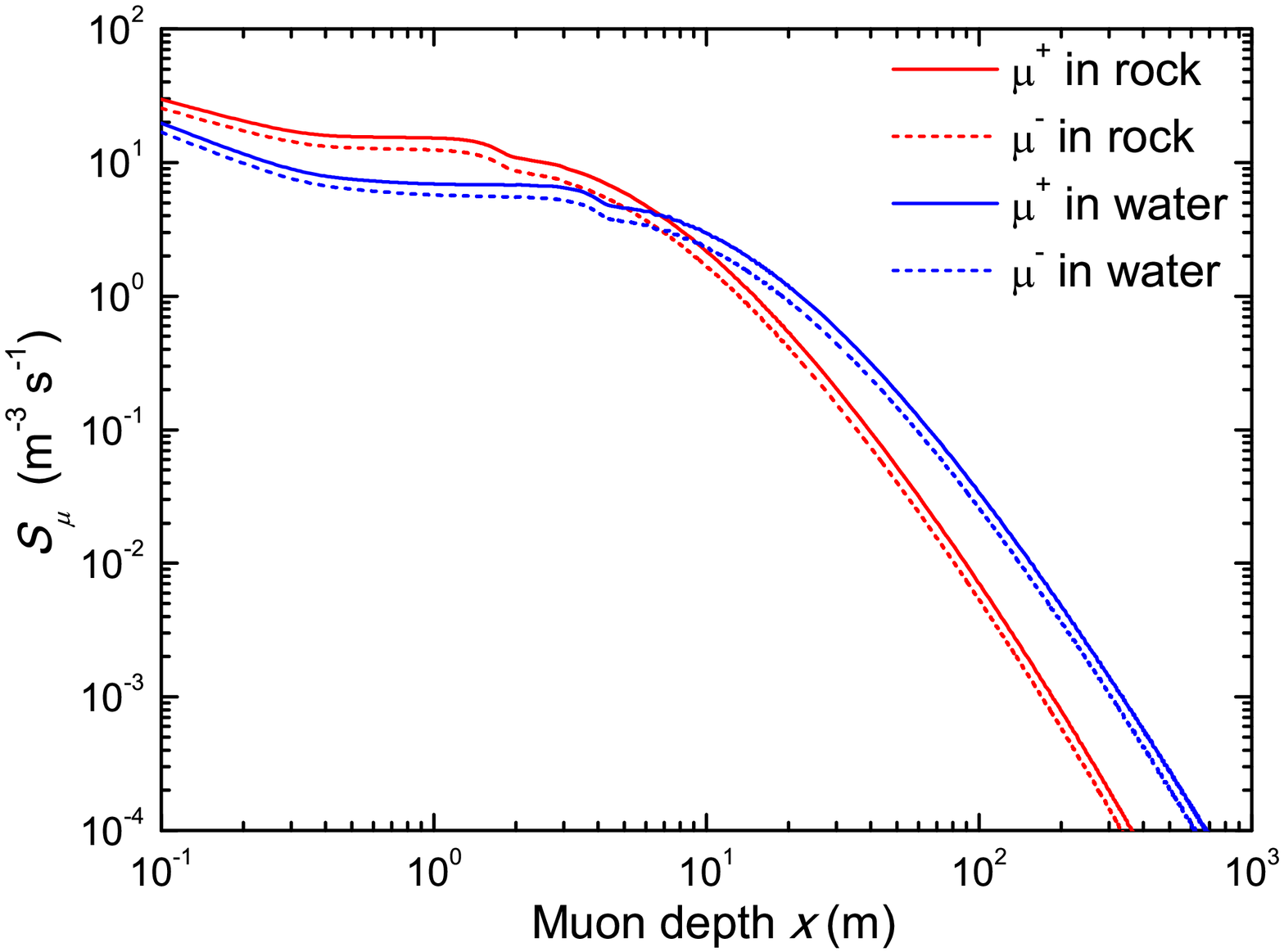}
\includegraphics[scale=0.38]{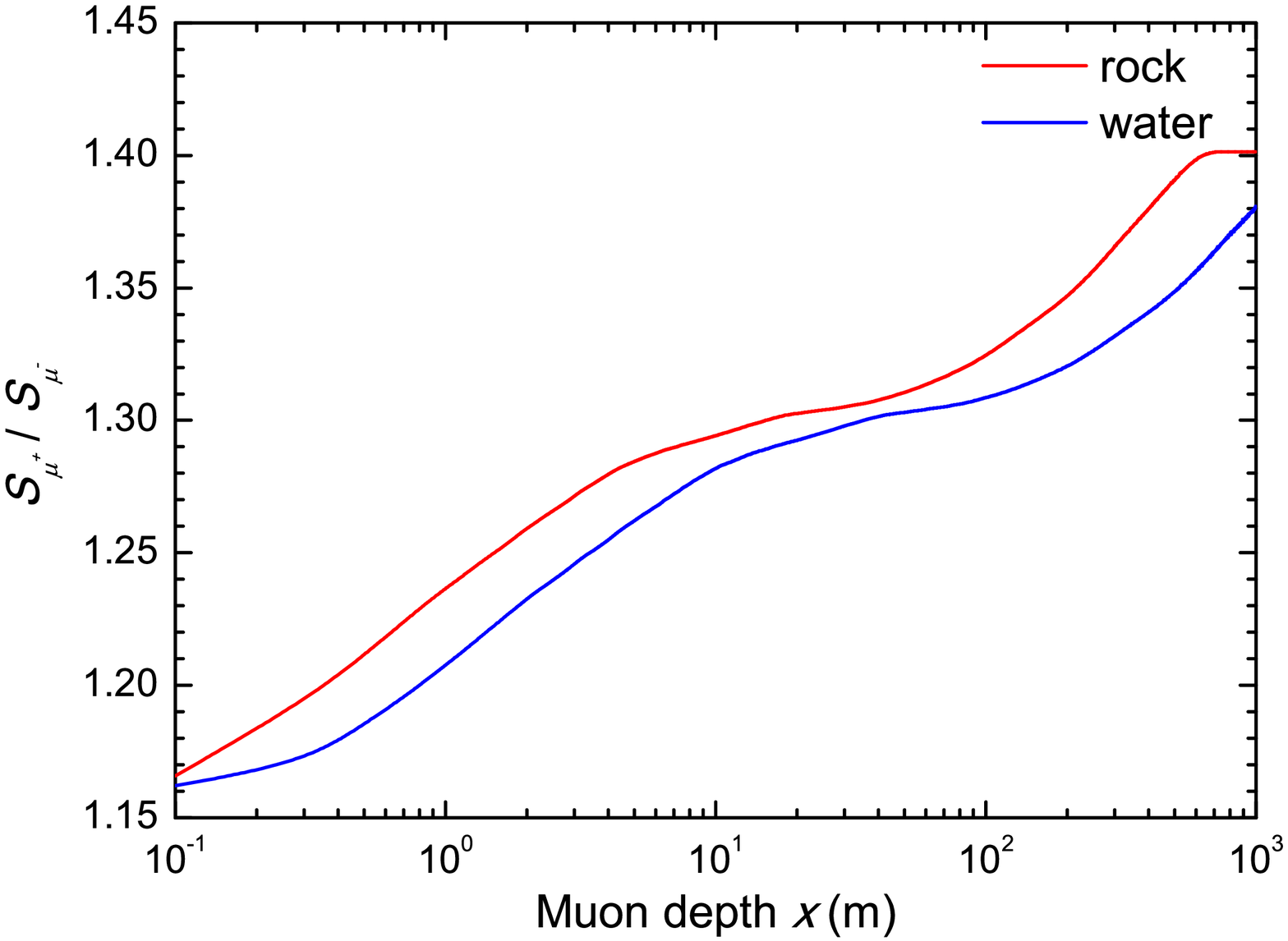}
\end{center}
\vspace{-1.0cm}\caption{ The muon stop rate per unit volume
$S_{\mu^\pm}$ (left) and the charge ratio of stopped muons
$S_{\mu^+}/S_{\mu^-}$ (right) as a function of stop depth $x$ in the
standard rock and water.} \label{MuonDist}
\end{figure}

The long-lived muons can penetrate the rock and sea water of the
Earth's surface to significant depths. The muon range $X(p_\mu)$ in
the standard rock and water may be found in Ref.
\cite{Groom:2001kq}. Then one can easily get the stop depth $x =
X(p_\mu) \cos \theta$ for a muon with the momentum $p_\mu$ and
incidence zenith angle $\theta$ at the sea level. Based on the muon
distribution $I(p_\mu , \theta)$, muon charge ratio
\cite{Naumov:2002dm} and muon range $X(p_\mu)$ \cite{Groom:2001kq},
we calculate the $\mu^\pm$ stop rate per unit volume
$S_{\mu^\pm}(x)$ in the standard rock and water. In Fig.
\ref{MuonDist}, the $\mu^\pm$ stop rate $S_{\mu^\pm}(x)$ and the
charge ratio of stopped muons $S_{\mu^+}/S_{\mu^-}$  as a function
of stop depth $x$  have been plotted. It is clear that the depth $x$
of most stopped muons is less than 30 m.

\section{Neutrino energy spectra from a stopped muon  \label{S2}}

It is well known that a stopped $\mu^+$  will quickly decay into a
positron and two neutrinos through $\mu^+ \rightarrow e^+ + \nu_e +
\bar{\nu}_\mu$. The $\nu_e$ and $\bar{\nu}_\mu$ energy spectra
(normalized to 1) can be written as \cite{Coloma:2017egw}
\begin{eqnarray}
f_{\nu_e} & = & \frac{192}{m_\mu} \left[ \left( \frac{E_\nu}{m_\mu}
\right)^2 \left( \frac{1}{2} - \frac{E_\nu}{m_\mu}  \right) \right]
\;, \label{f1} \\
f_{\bar{\nu}_\mu} & = & \frac{64}{m_\mu} \left[ \left(
\frac{E_\nu}{m_\mu} \right)^2 \left( \frac{3}{4} -
\frac{E_\nu}{m_\mu}  \right) \right] \;,  \label{f2}
\end{eqnarray}
where $m_\mu$ is the muon mass and $E_\nu \leq m_\mu/2$. Unlike
$\mu^+$, a stopped $\mu^-$ can not only decay $\mu^- \rightarrow e^-
+ \bar{\nu}_e +\nu_\mu$, but also be captured by nucleus and produce
a neutrino $\nu_\mu$ with $E_\nu < m_\mu$, such as $\mu^- +
^{16}{\rm O} \rightarrow ^{16}{\rm N}^\ast + \nu_\mu$. In fact, the
stopped $\mu^-$ will be quickly attached to an atom and form a
muonic atom (Atomic capture) \cite{Measday:2001yr} when it stops in
the rock or sea water. Then it cascades down to the lowest 1$s$
level in a time-scale of the order of $10^{-13}$ s through emitting
Auger electrons and muonic X-rays. In the following time, the
bounded $\mu^-$ in a muonic atom has only two choices, to decay or
to capture on the nucleus (Nuclear capture) \cite{Measday:2001yr}.

In order to estimate the $\mu^-$ decay and nuclear capture
probabilities, we should firstly consider the relative abilities of
atomic capture for different elements in the rock. Egidy and
Hartmann \cite{VonEgidy:1982pe} find a semi-empirical approach and
give the average atomic capture probability $P(Z)$ for 65 elements,
normalized to 1 for $^{16}{\rm O}$. For the rock chemical
composition, we take the upper continental crust data from Ref.
\cite{Rundick}. Then the mass and number percentages of 10 dominant
elements in the upper continental crust have been calculated and
listed in Table \ref{DecayRate}. Considering the corresponding
atomic capture probability $P(Z)$, we derive the atomic capture
percentages of 10 elements as shown in the fifth column of Table
\ref{DecayRate}. It is worthwhile to stress that the water is a
$^{16}{\rm O}$ target since $\mu^- p$ can easily penetrate nearby
 $^{16}{\rm O}$ atoms \cite{Measday:2001yr}.

\begin{table}[htb]
\begin{center}
\begin{tabular}{|c|c|c|c|c||c|c|c|}
\hline \hline
 Elements & \; Mass ($\%$) \; & \; Number ($\%$) \; & \; $P(Z)$ \; & Atomic capture ($\%$) & $\tau_{\mu^-}$ (ns) & Huff factor & $D_{\mu^-}$ ($\%$) \\ \hline
 O  & 47.51 & 62.13 & 1.00 & 60.26 & 1795.4 & 0.998 & 81.56 \\ \hline
 Si & 31.13 & 23.89 & 0.84 & 19.46 & 756    & 0.992 & 34.14 \\ \hline
 Al & 8.15  & 3.91  & 0.76 & 2.88  & 864    & 0.993 & 39.05 \\ \hline
 Fe & 3.92  & 2.27  & 3.28 & 7.21  & 206    & 0.975 & 9.14  \\ \hline
 Ca & 2.57  & 2.07  & 1.90 & 3.81  & 332.7  & 0.985 & 14.92 \\ \hline
 Na & 2.43  & 2.27  & 1.00 & 2.21  & 1204   & 0.996 & 54.58 \\ \hline
 K  & 2.32  & 1.28  & 1.54 & 1.91  & 435    & 0.987 & 19.54 \\ \hline
 Mg & 1.50  & 1.99  & 0.93 & 1.79  & 1067.2 & 0.995 & 48.33 \\ \hline
 Ti & 0.38  & 0.17  & 2.66 & 0.45  & 329.3  & 0.981 & 14.70 \\ \hline
 P  & 0.07  & 0.02  & 1.04 & 0.02  & 611.2  & 0.991 & 27.57 \\ \hline
\hline
\end{tabular}
\end{center}
\vspace{-0.2cm} \caption{The $\mu^-$ atomic capture percentages and
decay probabilities $D_{\mu^-}$ in a muonic atom for 10 dominant
elements of the upper continental crust \cite{Rundick}. The
corresponding mass and number percentages, average atomic capture
probability $P(Z)$, $\mu^-$ mean life $\tau_{\mu^-}$ and Huff factor
$Q$ have also been listed.} \label{DecayRate}
\end{table}

The decay rate $\Lambda_{\rm decay}$ and nuclear capture rate
$\Lambda_{\rm capture}$ of the bounded $\mu^-$ in a muonic atom have
the following relation \cite{Measday:2001yr}
\begin{eqnarray}
\Lambda_{\rm total} = \Lambda_{\rm capture} + Q \Lambda_{\rm decay}
\,,
\end{eqnarray}
where $\Lambda_{\rm total} = 1/\tau_{\mu^-}$, $\Lambda_{\rm decay} =
1/\tau_{\mu^+}$, and $Q$ is the Huff factor \cite{Suzuki:1987jf}.
Then the $\mu^-$ decay probability can be easily obtained by
\begin{eqnarray}
D_{\mu^-} = Q \frac{\Lambda_{\rm decay}}{\Lambda_{\rm total}} =  Q
\frac{\tau_{\mu^-}}{\tau_{\mu^+}} \,.
\end{eqnarray}
With the help of $\tau_{\mu^+} = 2196.98$ ns \cite{PDG} and the
$\mu^-$ mean life in Ref. \cite{Suzuki:1987jf},  we calculate the
$\mu^-$ decay probabilities $D_{\mu^-}$ for 10 dominant elements as
listed in the last column of Table \ref{DecayRate}. Combining the
atomic capture percentages and the corresponding $D_{\mu^-}$ in
Table \ref{DecayRate}, one can find that the averaged decay
probability $D_{\mu^-} =60.65\%$ and nuclear capture probability
$C_{\mu^-} =39.35\%$ for negative muons stopped in the rock. Since
the water can be approximated as an $^{16}{\rm O}$ target, the
$\mu^-$ decay and nuclear capture probabilities in the water are
$D_{\mu^-} = 81.56\%$ and $C_{\mu^-} = 18.44\%$, respectively. For
the $\bar{\nu}_e$ and $\nu_\mu$ energy spectra, we ignore the
differences between the free $\mu^-$ decay and the bounded $\mu^-$
decay \cite{Measday:2001yr}, and have
\begin{eqnarray}
f_{\bar{\nu}_e} & = &  f_{{\nu}_e} D_{\mu^-} \;, \label{f3}\\
f_{{\nu}_\mu} & = & f_{\bar{\nu}_\mu} D_{\mu^-} +
\tilde{f}_{\nu_\mu} C_{\mu^-} \;, \label{f4}
\end{eqnarray}
where $\tilde{f}_{\nu_\mu}$ is the $\nu_\mu$ energy spectrum
(normalized to 1) from the $\mu^-$ nuclear capture. It is found that
$\tilde{f}_{\nu_\mu}$ is fairly similar to the $\gamma$ spectrum in
the reaction of the $\pi^-$ capture on nucleus
\cite{Measday:2001yr}. Therefore we use the  $\gamma$ spectrum from
the $^{16}{\rm O} (\pi^-, \gamma ) ^{16}{\rm N}^*$ experimental
results \cite{Strassner:1979pz} and require that the maximal
neutrino energy only reaches 95 MeV for $\tilde{f}_{\nu_\mu}$,
because the  muon mass is 34 MeV less than that of a pion.

\section{Neutrino fluxes from stopped muons}

\begin{figure}[htb]
\begin{center}
\includegraphics[scale=0.38]{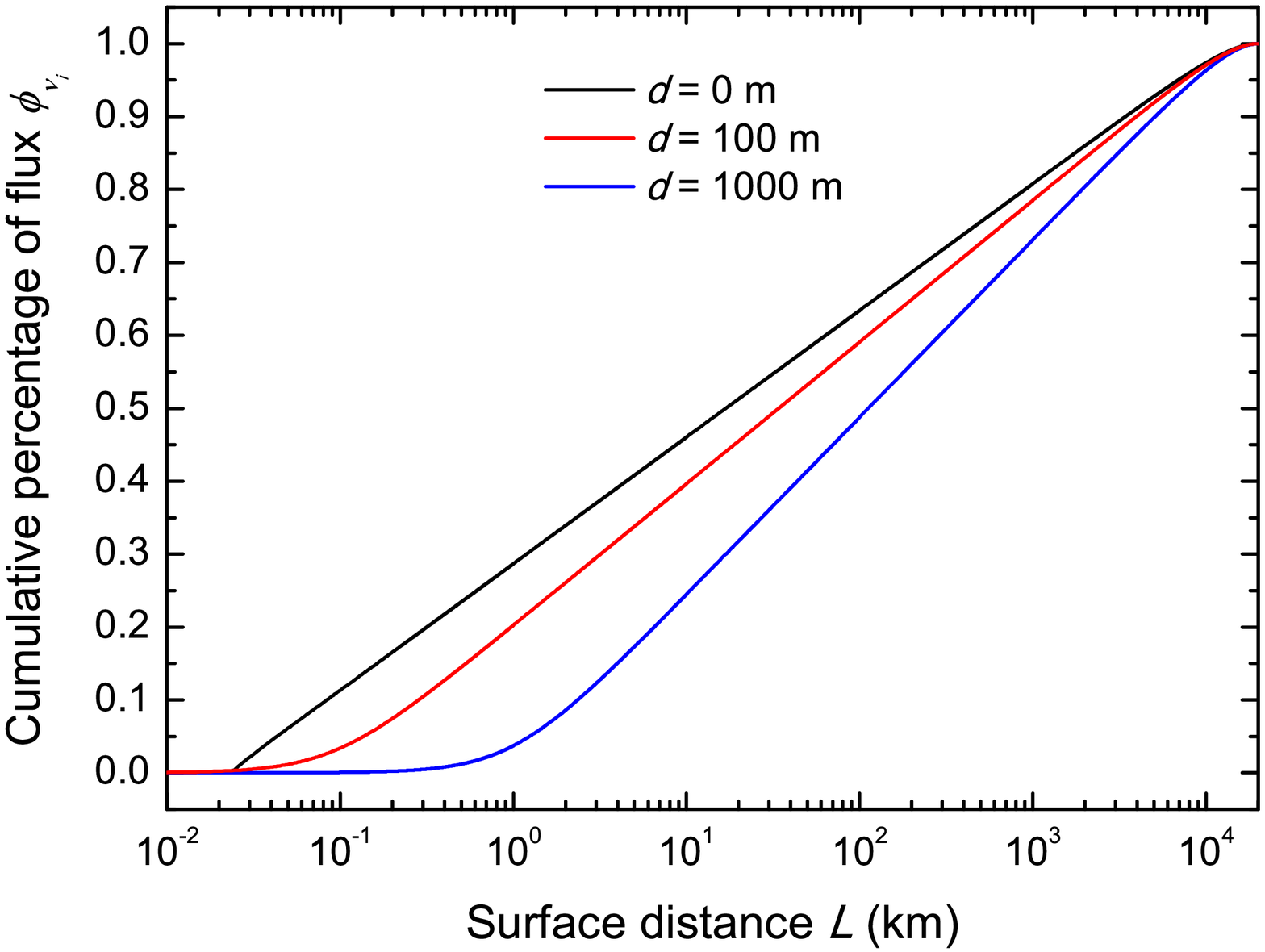}
\includegraphics[scale=0.38]{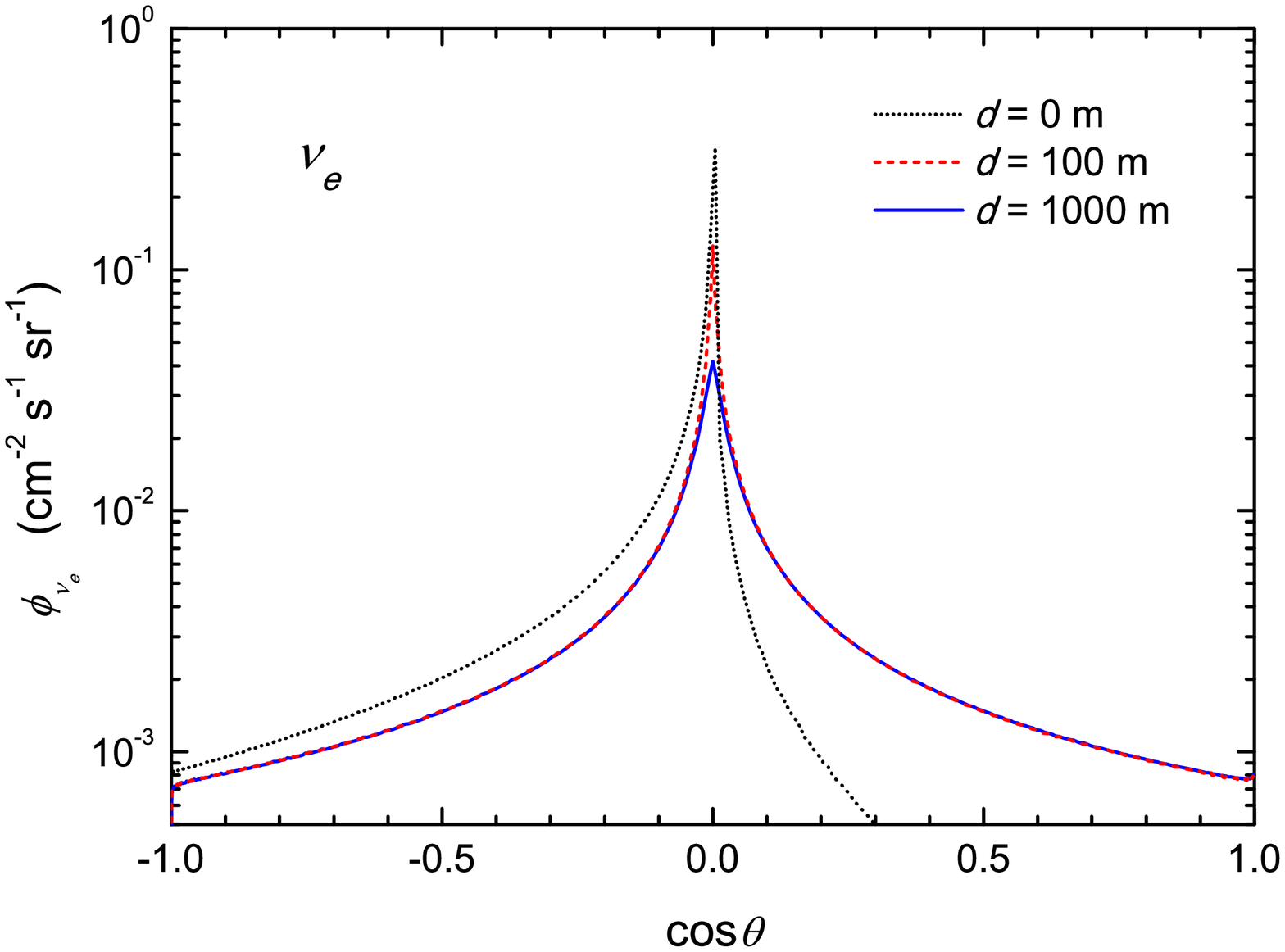}
\end{center}
\vspace{-1.0cm}\caption{The cumulative percentage of the neutrino
integral flux $\phi_{\nu_i}$ as a function of surface distance $L =
R_\oplus \vartheta$ (left) and the $\nu_e$ integral flux
$\phi_{\nu_e}$ with $E_\nu < 53$ MeV as a function of zenith angle
$\theta$ (right).} \label{LandAngular}
\end{figure}

Since the produced neutrinos from stopped muons are isotropic, the
neutrino differential fluxes can be written as
\begin{eqnarray}
\frac{d \phi_{\nu_i}}{d E_{\nu}} & = & f_{\nu_i} \int S_{\mu^\pm}(x)
\frac{ 2 \pi (R_\oplus -x)^2 \sin \vartheta }{4 \pi r^2} d \vartheta
d x \,, \label{FluxEq}
\end{eqnarray}
where $R_\oplus =6371$ km is the Earth's radius, and $\vartheta$ is
the angle between the stopped $\mu^\pm$ and detector point seen from
the Earth's center. The distance between the stopped $\mu^\pm$ and
detector is given by
\begin{eqnarray}
r^2 = (R_\oplus -x)^2 + (R_\oplus -d)^2 - 2 \cos \vartheta (R_\oplus
-x)(R_\oplus -d) \,,
\end{eqnarray}
where $d$ is the detector depth. In fact, the neutrino oscillation
should be considered \cite{Peres:2009xe}, but it is beyond the scope
of this paper. For a large neutrino detector, different detector
parts will receive notably different neutrino fluxes from stopped
muons within $r<100$ m. Therefore we take a virtual spherical
detector with a 25 m radius for the following analysis, which will
increase about $1\%$ flux for the $d<30$ m case. The integral flux
$\phi_{\nu_i}$ can be obtained from Eq. (\ref{FluxEq}). In the left
panel of Fig. \ref{LandAngular}, we plot the cumulative percentage
of $\phi_{\nu_i}$ as a function of surface distance $L = R_\oplus
\vartheta$ for three typical detector depths in the rock case.
Different flavors have almost identical results even in the water
case. It is worthwhile to stress that the stopped muons within the
surface distance $L = 200$ km contribute $68.7 \%$ and $56.1\%$ of
$\phi_{\nu_i}$ for $d=0$ m and $d=1000$ m, respectively. In
addition, one may easily find that most neutrinos come from the near
horizontal direction. Since the recoil electrons can carry the
directional information of incident neutrinos in the elastic
scattering of neutrinos on electrons \cite{Ueno:2012md} and $\nu_e$
has the largest cross section, we calculate the zenith angular
distribution of $\nu_e$ integral flux in the rock case, as shown in
the right panel of Fig. \ref{LandAngular}. It is found that the
smaller $d$ case has the narrower peak at the horizontal direction.
Note that the underground part of the virtual spherical detector
contributes the $\cos \theta >0$ angular distribution for the depth
of detector center $d =0$ m case. $\bar{\nu}_e$, ${\nu}_\mu$ and
$\bar{\nu}_\mu$ have similar angular distributions even in the water
case.

\begin{figure}[htb]
\begin{center}
\includegraphics[scale=0.38]{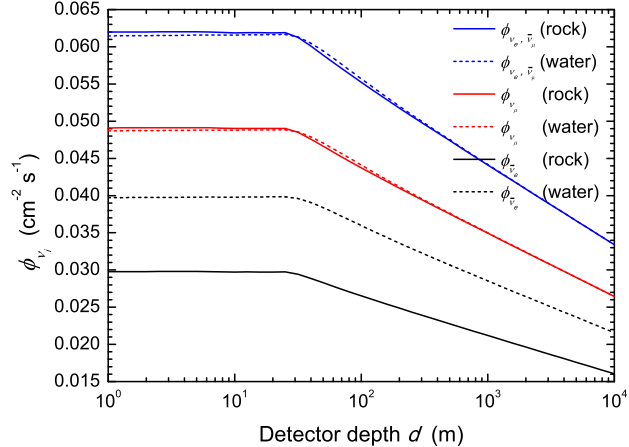}
\end{center}
\vspace{-1.0cm}\caption{The neutrino flux $\phi_{\nu_i}$ as a
function of detector depth $d$ for different neutrino flavors.}
\label{Fluxes}
\end{figure}

In Fig. \ref{Fluxes}, we show the neutrino integral fluxes
$\phi_{\nu_i}$ as a function of detector depth $d$ for different
flavors in the rock and water cases. It is clear that $\phi_{\nu_i}$
depends heavily on the detector depth $d$. For $\phi_{\nu_e}$,
$\phi_{\bar{\nu}_\mu}$ and $\phi_{{\nu}_\mu}$, the differences
between the rock and water cases are very small. However, the
${\nu}_\mu$ differential fluxes in the rock and water cases have
obvious differences due to different values of $D_{\mu^-}$ and
$C_{\mu^-}$ in Eq. (\ref{f4}). Note that $\phi_{\bar{\nu}_e}$ in the
water case is larger than that in the rock case because of the
larger $D_{\mu^-}$ in the water case. Therefore, the $\bar{\nu}_e$
flux depends on the local rock and water distributions for a given
detector.

Before presenting the differential neutrino fluxes, we here
introduce an approximation formula to calculate the neutrino fluxes.
Since the depth of most stopped muons is less than 30 m as shown in
Fig. \ref{MuonDist}, one may assume $S_{\mu^\pm}(x) = J_{\mu^\pm} \,
\delta (x)$ and subsequently simplify Eq. (\ref{FluxEq}) to
\begin{eqnarray}
\frac{d \phi_{\nu_i}}{d E_{\nu}}  & = & f_{\nu_i} J_{\mu^\pm} A(d)
\,, \label{Appro1}
\end{eqnarray}
where $A(d) = 0.5 R_\oplus^2  \int \sin \vartheta r_{x=0}^{-2}  d
\vartheta$ expresses the conversion factor from the muon flux
$J_{\mu^\pm}$ at the sea level to the neutrino flux $\phi_{\nu_i}$
for a detector with a depth $d$, and $A(R_\oplus) = 1$. For $d <
10^4$ m, $A(d)$ can be approximated as
\begin{eqnarray}
A(d) = {\rm Min} [ 6.62, 8.39-1.21 \log d ]\,, \label{Appro2}
\end{eqnarray}
where $d$ is in units of meter. It is found that Eqs.(\ref{Appro1})
and (\ref{Appro2}) can describe the exact results shown in Fig.
\ref{Fluxes} with an error of $2\%$.

\begin{figure}[htb]
\begin{center}
\includegraphics[scale=0.38]{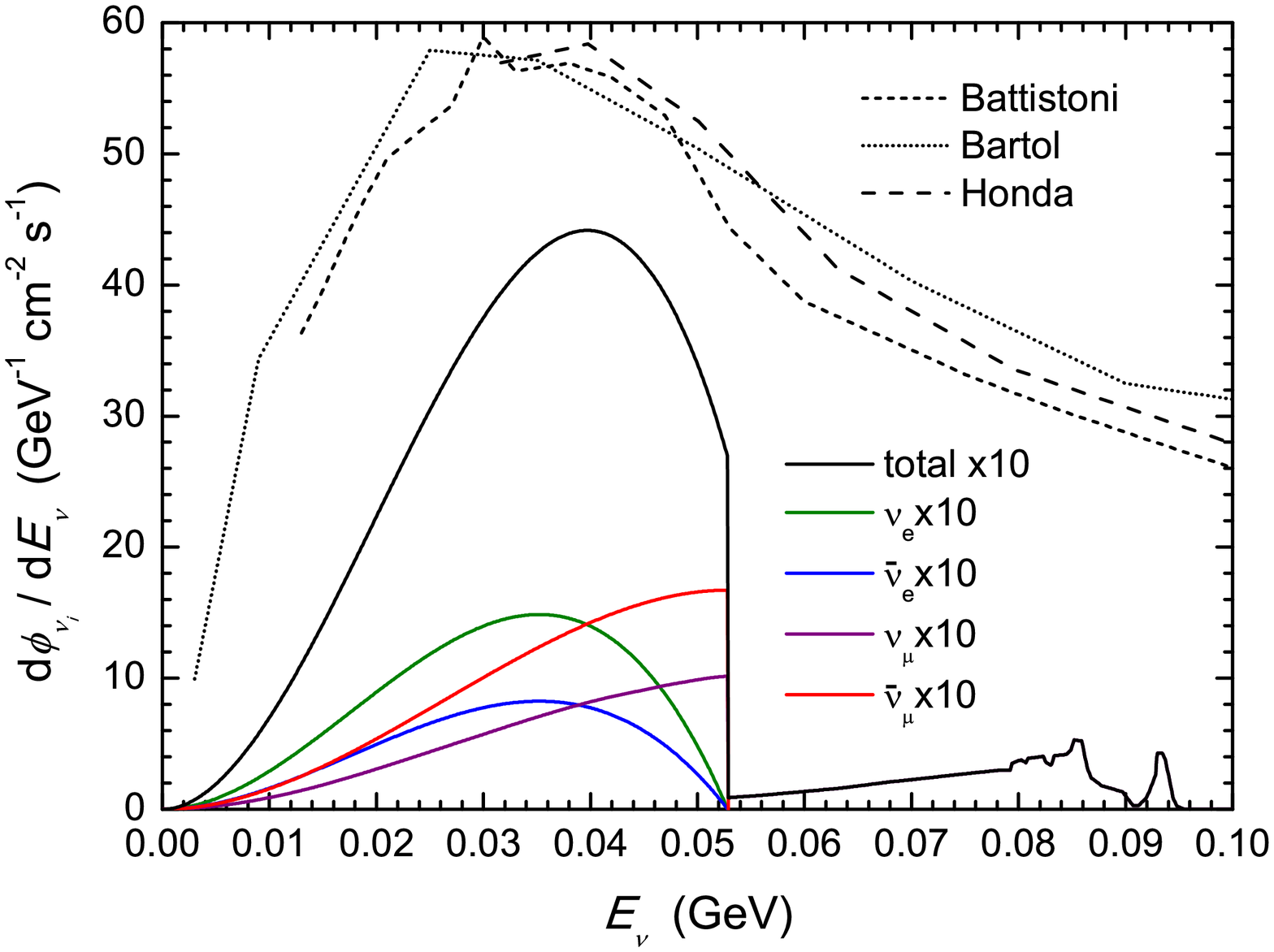}
\includegraphics[scale=0.38]{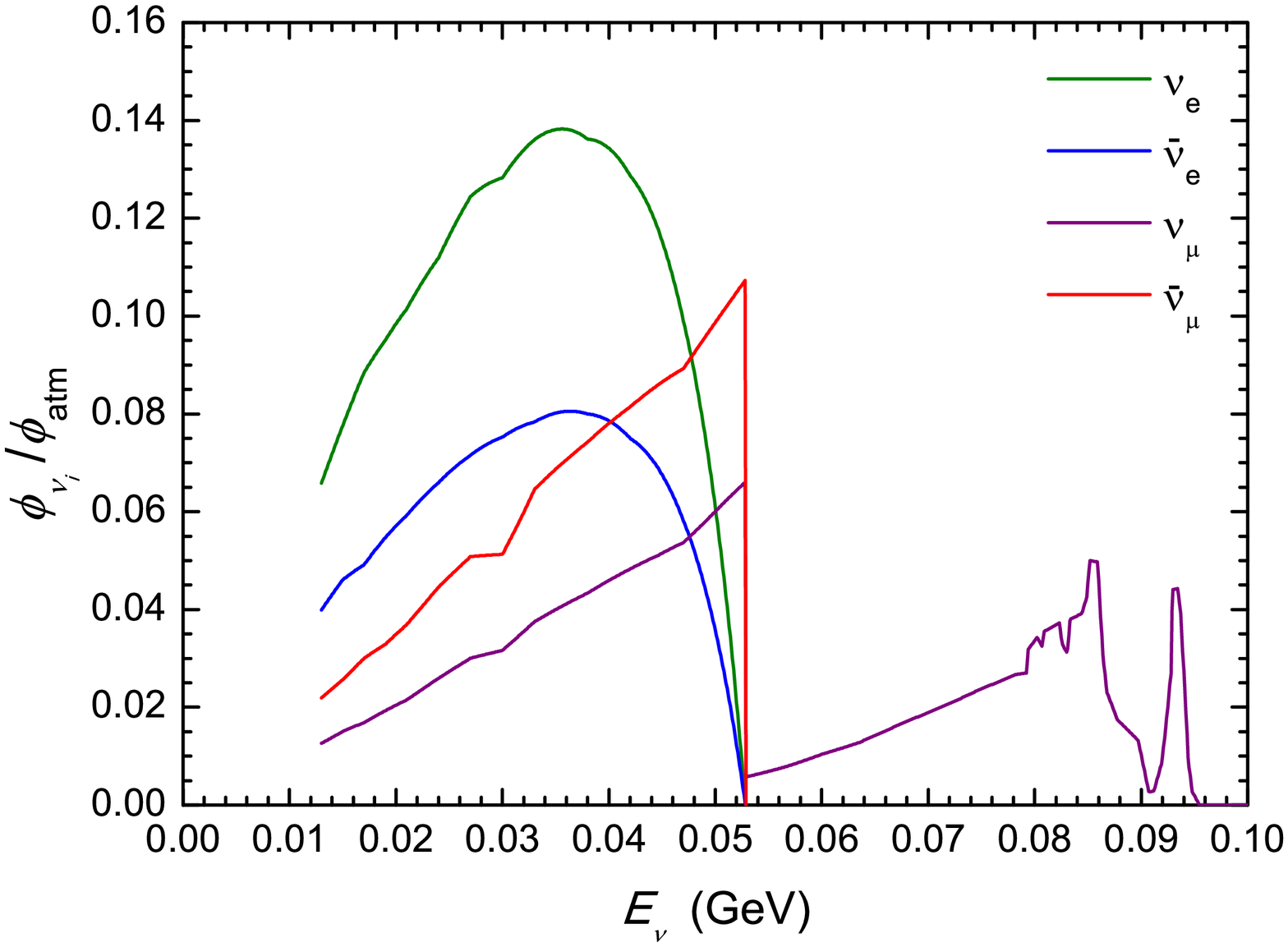}
\end{center}
\vspace{-1.0cm}\caption{ The differential neutrino fluxes $d
\phi_{\nu_i}/ d E_{\nu}$ (left) and the ratios of $d \phi_{\nu_i}/ d
E_{\nu}$ to the corresponding atmospheric neutrino flux from Ref.
\cite{Battistoni:2005pd} (right) for different flavors and $d=1000$
m. In the left panel, the total atmospheric neutrino fluxes at the
Super-K site from the Bartol \cite{Gaisser:1988ar}, Honda
\cite{Honda:1995hz} and Battistoni \cite{Battistoni:2005pd} groups
have also been shown.} \label{diffFlux}
\end{figure}

By use of Eq. (\ref{FluxEq}), we calculate the differential neutrino
fluxes as shown in the left panel of Fig. \ref{diffFlux}. Here
$D_{\mu^-} =70\%$ ($C_{\mu^-} =30\%$) and a Super-K detector depth
$d = 1000$ m have been assumed. For comparison,  the total
atmospheric neutrino fluxes at the Super-K site from the Bartol
\cite{Gaisser:1988ar}, Honda \cite{Honda:1995hz} and Battistoni
\cite{Battistoni:2005pd} groups have also been shown. It is found
that the neutrino fluxes from stopped muons are much less than the
atmospheric neutrino fluxes. In the right panel of Fig.
\ref{diffFlux}, we plot the ratios of $\phi_{\nu_i}$ to the
corresponding atmospheric neutrino flux from the Battistoni group
\cite{Battistoni:2005pd} for different flavors. For 13 MeV $ \leq
E_\nu \leq$ 53 MeV, the $\nu_e$, $\bar{\nu}_e$, $\nu_\mu$ and
$\bar{\nu}_\mu$ fluxes are averagely  $10.8 \%$, $6.3\%$, $3.7 \%$
and $6.2 \%$ of the corresponding atmospheric neutrino fluxes,
respectively. It is worthwhile to stress that the above results will
be increased by a factor of 1.4 if the detector depth $d < 30$ m.

\section{Conclusions}

In conclusion, we have investigated the low energy neutrinos from
stopped cosmic ray muons in the Earth. The $\mu^\pm$ stop rates per
unit volume $S_{\mu^\pm}(x)$ in the rock and sea water have been
calculated in terms of the muon intensity $I(p_\mu , \theta)$ at the
sea level and the muon range $X(p_\mu)$. Based on the atomic capture
and nuclear capture abilities of 10 dominant elements in the upper
continental crust, we estimate the $\mu^-$ decay and nuclear capture
probabilities in the rock and sea water. Then the neutrino energy
spectra $f_{\nu_e}$, $f_{\bar{\nu}_e}$, $f_{\nu_\mu}$ and
$f_{\bar{\nu}_\mu}$ from a stopped muon are given. Finally, we
present the low energy neutrino fluxes and give simultaneously a
good approximation to calculate them. It is found that most
neutrinos come from the surface distance $L < 200$ km region and the
near horizontal direction. For the integral fluxes $\phi_{\nu_e}$,
$\phi_{\bar{\nu}_\mu}$ and $\phi_{{\nu}_\mu}$, the differences
between the rock and water cases are very small. On the contrary,
the $\phi_{\bar{\nu}_e}$ depends on the local rock and water
distributions because of different $\mu^-$ decay probabilities. Note
that all $\phi_{\nu_i}$ depend heavily on the detector depth $d$.
For the Super-K detector depth $d = 1000$ m, the $\nu_e$,
$\bar{\nu}_e$, $\nu_\mu$ and $\bar{\nu}_\mu$ fluxes in the range of
13 MeV $ \leq E_\nu \leq$ 53 MeV are averagely  $10.8 \%$, $6.3\%$,
$3.7 \%$ and $6.2 \%$ of the corresponding atmospheric neutrino
fluxes, respectively. The above results will be increased by a
factor of 1.4 if the detector depth $d < 30$ m. These low energy
neutrinos should be considered in searches of some related topics.

\acknowledgments

We are grateful to Meng-Yun Guan and Ji-Lei Xu for their useful
discussions and helps. This work is supported in part by the
National Nature Science Foundation of China (NSFC) under Grants No.
11575201 and No. 11835013, and the Strategic Priority Research
Program of the Chinese Academy of Sciences under Grant No.
XDA10010100.


\end{document}